\documentclass{IEEEcsmag}

\usepackage[colorlinks,urlcolor=blue,linkcolor=blue,citecolor=blue]{hyperref}

\usepackage{upmath,amsmath,amsfonts}

\usepackage{algorithm2e,xcolor}
\newcommand{\be}{\begin{equation}}
\newcommand{\ee}{\end{equation}}
\newcommand{\bea}{\begin{eqnarray}}
\newcommand{\eea}{\end{eqnarray}}
\newcommand{\bes}{\begin{equation*}}
\newcommand{\ees}{\end{equation*}}
\newcommand{\beas}{\begin{eqnarray*}}
\newcommand{\eeas}{\end{eqnarray*}}

\newcommand{\tr}{\mathrm{tr}}

\begin{document}


\title{ Measurement Crosstalk Errors in Cloud-Based Quantum Computing}

\author{Seungchan Seo}
\affil{School of Electrical Engineering, Korea Advanced Institute of Science and Technology (KAIST), 291 Daehak-ro, Yuseong-gu, Daejeon 34141, Republic of Korea }

\author{Joonwoo Bae}
 \affil{School of Electrical Engineering, Korea Advanced Institute of Science and Technology (KAIST), 291 Daehak-ro, Yuseong-gu, Daejeon 34141, Republic of Korea }


\begin{abstract}
Quantum technologies available currently contain noise in general, often dubbed noisy intermediate-scale quantum (NISQ) systems. We here present the verification of noise in measurement readout errors in cloud-based quantum computing services, IBMQ and Rigetti, by directly performing quantum detector tomography, and show that there exist measurement crosstalk errors. We provide the characterization and the quantification of noise in a quantum measurement of multiple qubits. We remark that entanglement is found as a source of crosstalk errors in a measurement of three qubits. 
 \end{abstract}

\maketitle

\chapterinitial{The computation} based on the laws of quantum mechanics makes it possible to achieve the capability beyond the limitations of conventional computing \cite{AA1,BB1,CC1}. To understand the existing gap between theory and implementation in the realization of quantum computing, the key is to find how noise deteriorates the capability of quantum information processing. Noise in a quantum system signifies its transition to a classical one. Once systems are governed by classical physics, no quantum advantages can be obtained. An ultimate solution to deal with quantum noise may be obtained by quantum error-correcting codes that can preserve quantum states \cite{GG1}. The quantum technologies available at present, dubbed noisy intermediate-scale quantum (NISQ) technologies \cite{DD1}, do not meet the level to implement quantum error correction.

Although the errors are present, NISQ technologies can be used to show advantages over classical systems \cite{google}. Much effort has been devoted to improving the capability of NISQ-based information processing. Quantum algorithms fitted to NISQ technologies are devised, e.g., quantum approximate optimization \cite{EE1} or variational quantum eigensolvers \cite{FF1}. Strategies to mitigate the errors in quantum dynamics have been proposed while quantum error-correcting codes cannot be realized \cite{endo}, see also \cite{em1,em2,em3}. In fact, industry vendors such as IBMQ, IonQ, or Rigetti immediately provide cloud-based quantum computing services. 

\begin{figure*}
\centerline{\includegraphics[width=15.5cm]{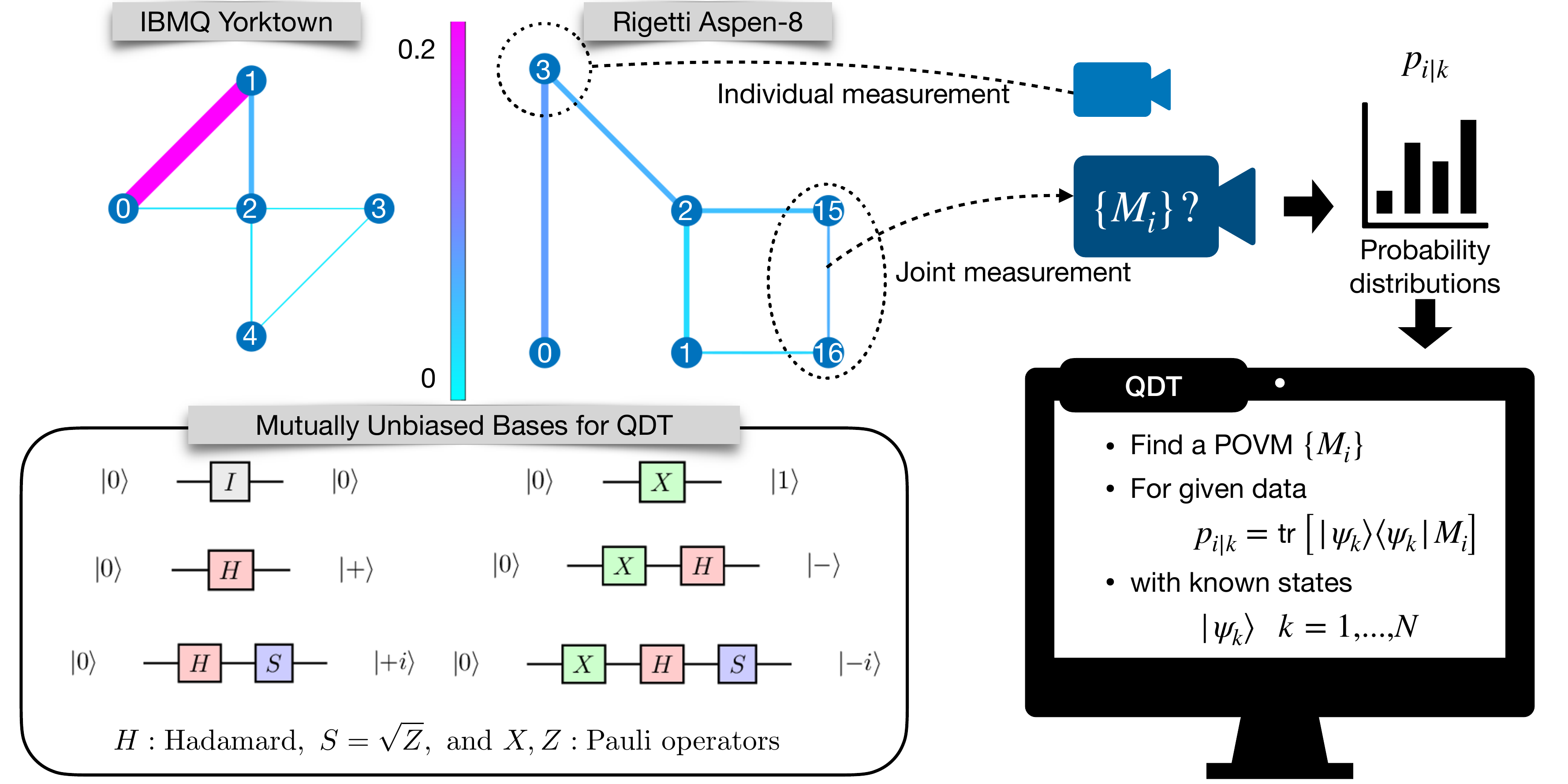}}
\caption{  QDT is performed on detectors in IBMQ Yorktown and Rigetti Aspen-8. Qubit states are prepared in the six states and measured. QDT finds measurement operators from the statistics of measurement outcomes. }
\label{scenario}
\end{figure*}

The main focus of the present article is measurement readout errors, in particular, crosstalk errors, appearing in a quantum measurement with NISQ technologies. The significance of measurement errors is twofold. Firstly, no method is known so far to generally clear measurement errors. Quantum error-correcting codes that aim to deal with noise appearing in quantum dynamics cannot apply to measurement readout errors. Remarkably, readout errors on individual qubits can be mitigated with the help of quantum detector tomography (QDT) \cite{HH1}. The step of QDT that costs expensive experimental resources may be circumvented by depolarizing various types of quantum noise with the help of single-qubit gates \cite{II1}. Secondly, evidences for crosstalk errors beyond individual qubits have been reported \cite{JJ1}. For instance, crosstalk among detectors may cause unexpected correlations in the outcome statistics. This means that one should not consider measurement errors on individual qubits only. Moreover, it is generally hard to verify the source for measurement crosstalk errors, which may be caused by interactions within systems but an environment. All these make it even harder to deal with measurement errors in general.

In this work, we show that crosstalk errors in a measurement of two and three qubits in fact exist in the cloud-based quantum computing services, IBMQ and Rigetti, by performing QDT directly on the detectors for multiple qubits. In particular, entanglement is detected in the measurement operators describing three-qubit detectors but two-qubit ones. We introduce an operational method to quantify measurement crosstalk errors and provide the analysis. Our results can be used to devise methods to mitigate measurement errors.

The paper is organized as follows. We begin by a brief review of QDT. We then introduce the characterization and quantification of crosstalk errors in a quantum measurement. Measurement errors in IBMQ Yorktown and Rigetti Aspen-8 are analyzed. Finally, we summarize the structure of correlations appearing in a measurement of multiple qubits.

\section{ Quantum Detector Tomography }
\label{pre}

Let us begin with the measurement postulate in quantum theory. The quantum measurement dictates the statistics of measurement outcomes in a quantum experiment. A quantum probability is given by a product of a state and a measurement operator. A quantum measurement is generally described by positive-operator-valued-measures (POVMs), denoted by $\{M_i \}_{i}$, that satisfy
\bea
M_i\geq0,~\sum_{i}M_i = I. \label{eq:m}
\eea 
For a qubit state identified as a two-level quantum system, a qubit measurement is described by $2\times 2$ POVM elements. For a quantum state $\rho$, the probability of obtaining an outcome $i$ is given by 
\bea
p( i ) =\tr[ M_i \rho]\nonumber 
\eea 
which is called the Born rule.

\subsection{ Measurement in quantum computing}

In quantum computing, a measurement is performed in the computational basis, for which a two-outcome measurement contains two POVM elements,
\bea
M_0 = |0\rangle \langle 0|~~\mathrm{and}~~M_1 = |1\rangle \langle 1|. \label{eq:qm}
\eea
A detection event in the arm $0$ ($1$) is described by $M_0$ ($M_1$). A measurement for $n$ qubits giving outcomes $a_j \in \{ 0,1\}$ for $j=1,\cdots,n$ is described by a POVM element in the following,
\bea
M_{\vec{a}} = M_{a_1}\otimes M_{a_2} \otimes \cdots \otimes M_{a_n}. \label{eq:nqm} 
\eea
Note that for $n$ detectors for $n$ qubits are characterized by $2^n$ POVM elements, each of which corresponds to a $2^n \times 2^n$ non-negative matrix. 

QDT aims to verify POVM elements for $n$-qubit detectors. Thus, non-negative $2^n\times 2^n$ matrices are obtained by QDT. In practice, QDT can be implemented by applying a set of tomographically complete states, repeating a measurement, collecting the outcome statistics and characterizing a non-negative matrix that is the most consistent with outcome statistics. An instance of the set of tomographically complete states is mutually unbiased states,
\bea
|0\rangle, ~|1\rangle, ~|+\rangle, ~|-\rangle, ~|+i\rangle, ~|-i\rangle, ~\label{eq:mub}
\eea
where $|\pm\rangle = (|0\rangle \pm |1\rangle )/\sqrt{2}$ and $|\pm i \rangle = (|0\rangle \pm i |1\rangle )/\sqrt{2}$. For $n$ qubits, the number of tomographically complete states is $O(4^n)$. Then, we apply maximum likelihood estimation to reconstruct a POVM element consistent to the outcome statistics \cite{qse}, see also the QDT algorithm.

\RestyleAlgo{ruled}
\SetKwRepeat{Do}{do:}{while}
\begin{algorithm}[hbt!]
    \caption{QDT with maximum likelihood estimation}
    \label{alg:MLE}
    \KwData{Prepare states $\rho_k = |\psi_k\rangle\langle \psi_k|$ in Eq. (\ref{eq:mub}) and the empirical frequency of obtaining outcome $i$, $\{ f_{i,k}\}$.}   
    \KwResult{A POVM $\{M_i\}$ that best describes given data, i.e., which maximizes log-likelihood function,
    $$\log\mathcal{L}=\sum_{i} \sum_{k} {f_{i,k}}\log tr[M_i\rho_k].$$}
     Set a termination threshold $\epsilon$.\\
     Start with $M_i^0=I/D$ where $D$ denotes a dimension of a system.\\
    \DontPrintSemicolon
    \Do{~$\sum_i {\Vert M_i^t - M_i^{t+1}\Vert}_1\ge\epsilon$}{
$p_{i,k}~\gets\tr~[M_i^t\rho_k]$\\
\hfill{$M^t$ and $R^t$ denote $M$ and $R$ at the $t$-th iteration step}\\
$R_i^t~\gets~ 
\sum_k\frac{f_{i,k}}{p_{i,k}} \times $\\
~~~~~~~~~$\left( \sum_j\sum_{l,m}\frac{f_{j,l}}{p_{j,l}}\frac{f_{j,m}}{p_{j,m}}\rho_l M_j^t\rho_m \right)^{-\frac{1}{2}}\rho_k$\\
$M_i^{t+1}~\gets~R_i^t M_i^t R_i^{\dagger,t}$
 }
\end{algorithm}

\subsection{ Quantum measurement in a realistic scenario}
In a realistic scenario with the NISQ systems, a measurement may not be the form in Equations (\ref{eq:qm}) or (\ref{eq:nqm}). First, POVM elements giving outcomes $0$ and $1$ in Equation (\ref{eq:qm}) may not be rank-one. This is due to noise on individual detectors, which we identify as local noise. Then, POVM elements for detecting multiple qubits may not be factorized in a form in Equation (\ref{eq:nqm}), that is,
\bea
M_{\vec{a}} \neq M_{a_1}\otimes M_{a_2} \otimes \cdots \otimes M_{a_n}. \label{eq:crossm} 
\eea
The type of noise may be due to crosstalk of detectors \cite{KK1}, which we call crosstalk errors in a measurement. In the following, we devise distance measures to identify crosstalk errors in a quantum measurement.



\begin{center}
\begin{table*}
\caption{  Two-qubit crosstalk errors in IBMQ Yorktown}
\label{table}
\small
\begin{tabular*}{36.2pc}{ | p{20pt} | p{88pt} |p{88pt} |p{88 pt}  | p{88 pt} |}
\hline
Qubits &  00 : $(D_N, D_C, D_{L}^{*})$  & 01 : $(D_N, D_C, D_{L}^{*})$  &10 : $(D_N, D_C, D_{L}^{*})$ & 11 : $(D_N, D_C, D_{L}^{*})$ \\
\hline
\hline
	(1,2) & (0.1158,0.0219,0.1159) & (0.1210,0.0155,0.1216) & (0.1710,0.0462,0.1708) & (0.1732,0.0140,0.1722) \\
	(2,3) & (0.0866,0.0105,0.0854) & (0.0519,0.0112,0.0544) & (0.0793,0.0122,0.0774) & (0.0553,0.0136,0.0573) \\
	(0,3) & (0.3312,0.0351,0.3365) & (0.3062,0.0449,0.3050) & (0.2614,0.0469,0.2629) & (0.2331,0.0625,0.2285) \\
	(1,4) & (0.1879,0.0056,0.1882) & (0.1267,0.0105,0.1257) & (0.2018,0.0125,0.2017) & (0.1452,0.0117,0.1446) \\
	(0,1) & (0.3295,0.0208,0.3299) & (0.3896,0.0307,0.3915) & (0.2588,0.1606,0.2513) & (0.2259,0.0273,0.2223) \\
	(3,4) & (0.1996,0.0064,0.1994) & (0.1148,0.0142,0.1145) & (0.1309,0.0129,0.1320) & (0.1099,0.0095,0.1113) \\
\hline
\multicolumn{5}{@{}p{36.2pc}@{}} { QDT for two-qubit detectors in IBMQ Yorktown is performed on Nov. 18 2020. The distance measures $(D_N, D_C, D_{L}^{*})$ are computed for the four POVM elements giving outcomes $00$, $01$, $10$ and $11$.  }
\end{tabular*}
\label{tab4}
\end{table*}
\end{center}

\section{Detection of Crosstalk Errors in Quantum Measurements}

To analyze crosstalk errors in a quantum measurement, it is essential to perform QDT. We exploit the trace distance to quantify crosstalk errors existing in POVM elements. The trace distance between two normalized POVM elements $\widetilde{\Pi}_1$ and $\widetilde{\Pi}_2$ is denoted by 
\bea
D( \widetilde{\Pi}_1, \widetilde{\Pi}_2) = \frac{1}{2}\| \widetilde{\Pi}_1 - \widetilde{\Pi}_2 \|_1, \nonumber
\eea
where $\| \cdot\|_1$ denotes the trace norm. 


\subsection{ Two-qubit crosstalk errors }

Suppose that a POVM element $\widetilde{\Pi}_{ab}^{xy}$ over two qubits labelled by $x$ and $y$ giving outcomes $a$ and $b$, respectively, is obtained from QDT. We now introduce a distance measure for the quantification of noise in the POVM element. Throughout, we consider normalized POVM elements. The total error is quantified by $D_N$
\bea
D_{N} (a,b|x,y) &: =& D( \widetilde{\Pi}_{ab}^{xy} , |ab \rangle\langle ab| ), ~~~~\label{eq:n} 
\eea
where $a,b\in \{0,1\}$. This shows the distance of a POVM element from the ideal one $|a\rangle\langle a| \otimes |b\rangle \langle b|$. A non-zero distance $D_N >0 $ may be contributed by two sources of noise: one is the errors in individual detectors due to local noise, and the other is the errors that cannot be described by noise on individual detectors. The latter is identified as crosstalk errors. 

The sources of noise may be elucidated by introducing crosstalk $D_C$ and local errors $D_L$,
\bea
D_{C} (a,b|x,y) &: =& \min_{ \widetilde{\Pi}_{a}^{x}  \otimes  \widetilde{\Pi}_{b}^{y}}   D( \widetilde{\Pi}_{ab}^{xy} ,  \widetilde{\Pi}_{a}^{x} \otimes  \widetilde{ \Pi}_{b}^{y} ),~~ ~~\label{eq:ct} \\
D_{L}^{*}  (a,b|x,y) &: =& D( \widetilde{\Pi}_{a}^{x} \otimes  \widetilde{\Pi}_{b}^{y}  , | ab\rangle\langle ab| ) ~~~\label{eq:l}
\eea
where the normalized POVM element $\widetilde{\Pi}_{a}^{x} \otimes  \widetilde{\Pi}_{b}^{y} $ in Equation (\ref{eq:l}) corresponds to the optimal one in Equation (\ref{eq:ct}). The measure $D_C$ finds the minimal distance of a POVM element from POVM elements having no crosstalk noise. One can conclude that crosstalk noise exists if and only if $D_C$ is positive: 
\bea
\mathrm{crosstalk ~ in~ detectors}~~\iff ~~ D_C>0. \nonumber
\eea
Once $D_C$ is computed with optimal local POVM elements, the distance between the optimal local POVM elements and the ideal one is measured by $D_{L}^{*}$. From the property of distance measures, it holds that
\bea
D_N\leq D_C + D_{L}^* ~~\mathrm{or}~~ D_C \geq D_N  - D_{L}^* \nonumber
\eea
for a given set $\{a,b,x,y \}$.


\begin{center}
\begin{table*}
\caption{Three-qubit measurement crosstalk errors in IBMQ Yorktown}
\label{table}
\small
\begin{tabular*}{35.2 pc}{ | p{70pt} |p{30pt} |p{30pt} |p{30pt}  | p{30pt} | p{30pt} |p{30pt} |p{30pt} |p{30pt} | }
\hline
Partitions   &  000 & 100 & 010 & 110 & 001 & 101 & 011 & 111 \\
	\hline
	\hline
	0:1:2 & 0.0242 & 0.1399 & 0.0591 & 0.0405 & 0.0292  & 0.1350 & 0.0366 & 0.0261 \\
	0:(2,3) & 0.0225 & 0.1383 & 0.0376 & 0.0238 & 0.0260 & 0.1325 & 0.0332 & 0.0212 \\
	1:(2,0) & 0.0189 & 0.1352 & 0.0538 & 0.0366 & 0.0268 & 0.1340 & 0.0315 & 0.0229 \\
        2:(0,1) & 0.0219 & 0.0260 & 0.0537 & 0.0367 & 0.0273 & 0.0247 & 0.0333 & 0.0164 \\
\hline
\multicolumn{9}{@{}p{35.2pc}@{}} {Detectors of theree qubits labelled by $0$, $1$, and $2$ are analyzed. From POVM elements giving outcomes $100$ and $101$, it is found that crosstalk errors in bipartite splittings $0:1$ are dominant. As for $2: (0,1)$, crosstalk errors are about $1\%$. }
\end{tabular*}
\label{tab5}
\end{table*}
\end{center}

\subsection{ Multi-qubit crosstalk errors }

For multiple qubits, crosstalk errors can be analyzed by bipartite splittings over multiple qubits. Let us consider a bipartition analysis for three qubits and the analysis can be generalized to multiple qubits straightforwardly. Let $A$, $B$, and $C$ denote three qubits, and $\widetilde{\Pi}_{abc}^{ABC}$ a normalized POVM element giving outcome $a$, $b$, and $c$.

The total error may be estimated as the trace distance,
\bea
D_N ( \widetilde{\Pi}_{abc}^{ABC} )= D( \widetilde{\Pi}_{abc}^{ABC}, |abc \rangle\langle abc| ). ~\label{eq:3n} 
\eea
The bipartite crosstalk errors may be quantified in the bipartite splittings $A:BC$, $B:CA$, and $C:AB$. For instance, the crosstalk error in the bipartition $A:BC$ is quantified by
\bea
&&D_{C} (\widetilde{\Pi}_{abc}^{A:BC} ) := \nonumber\\
&& \min_{ \widetilde{\Pi}_{a}^{A}  \otimes  \widetilde{\Pi}_{bc}^{BC}}   D( \widetilde{\Pi}_{abc}^{ABC} ,  \widetilde{\Pi}_{a}^{A} \otimes  \widetilde{ \Pi}_{bc}^{BC} ). ~~~~~~~~~~~~\label{eq:ctabc1} 
\eea
Similarly, the crosstalk error in other bipartite splittings can be quantified by,
\bea
&&D_{C} (\widetilde{\Pi}_{bca}^{B:CA} ) := \nonumber\\
&& \min_{ \widetilde{\Pi}_{b}^{B}  \otimes  \widetilde{\Pi}_{ca}^{CA}}   D( \widetilde{\Pi}_{abc}^{ABC} ,  \widetilde{\Pi}_{b}^{B} \otimes  \widetilde{ \Pi}_{ca}^{CA} ). ~~~~~~~~~~~~\label{eq:ctabc2} \\
&\mathrm{and}&D_{C} (\widetilde{\Pi}_{abc}^{C:AB} ) := \nonumber\\
&& \min_{ \widetilde{\Pi}_{c}^{C}  \otimes  \widetilde{\Pi}_{ab}^{AB}}   D( \widetilde{\Pi}_{abc}^{ABC} ,  \widetilde{\Pi}_{c}^{C} \otimes  \widetilde{ \Pi}_{ab}^{AB} ). ~~~~~~~~~~~~\label{eq:ctabc3} 
\eea
Given a POVM element above, one can also define local errors as it is shown in Equation (\ref{eq:l}). 

The genuine multi-qubit crosstalk errors can be quantified as follows,
\bea
&& D_{C} (\widetilde{\Pi}_{abc}^{A:B:C} ) := \nonumber \\
&& \min_{ \widetilde{\Pi}_{a}^{A}  \otimes  \widetilde{\Pi}_{b}^{B} \otimes  \widetilde{\Pi}_{c}^{C} }   D( \widetilde{\Pi}_{abc}^{ABC} ,  \widetilde{\Pi}_{a}^{A} \otimes  \widetilde{ \Pi}_{b}^{B}\otimes  \widetilde{\Pi}_{c}^{C} ).~~~ ~~\label{eq:ctabc4} 
\eea
Note that $D_{C} (\widetilde{\Pi}_{abc}^{A:B:C} ) =0 $ implies that crosstalk errors in all bipartite splittings in Equations (\ref{eq:ctabc1}), (\ref{eq:ctabc2}), and (\ref{eq:ctabc3}) are zero. Thus, it holds that non-zero crosstalk errors in a bipartite splitting imply $D_{C} (\widetilde{\Pi}_{abc}^{A:B:C} ) >0 $.

In what follows, we apply the quantification of crosstalk errors in a quantum measurement to the detectors in cloud-based quantum computing services, IBMQ Yorktown and Rigetti Aspen-8.

\section{ Crosstalk Errors in IBMQ Yorktown}

We have performed QDT for detectors in IBMQ Yorktown on Nov. 15 2020. IBMQ Yorktown contains five qubits, in which the connectivity is shown in Figure \ref{scenario}. QDT is performed on two- and three-qubit detectors.

For two-qubit detectors, the pairs of qubits labelled $(1,2)$, $(2,3)$, $(0,3)$, $(1,4)$, $(0,1)$, and $(3, 4)$ are analyzed. To find crosstalk errors, we have computed the distance $(D_N, D_C, D_{L}^{*})$ in Equation (\ref{eq:ct}) for the POVM elements of the pairs. They are listed in Table \ref{tab4}. It is found that two-qubit errors $D_N$ vary from $5\%$ to $39\%$ depending on the pairs. Errors around $30\%$ are found for the pairs $(0,3)$ and $(0,1)$ for which the qubit labelled by $0$ is in common. The pairs without qubit $0$, such as $(2,3)$, show relatively lower errors. The total error $D_N$ is higher in the pairs $(1,2)$ and $(0,3)$. It is found that crosstalk errors are around $1$-$5\%$ whereas local errors are $10$-$30\%$.

For three-qubit detectors, the triple of qubits labelled $(0,1,2)$ is analyzed. We have performed QDT on Nov. 11 2020. In Table \ref{tab5}, crosstalk errors in all partitions are computed. It is shown that the genuine crosstalk errors exist and peak at two POVM elements giving outcomes $100$ and $101$. Similar values are observed in the bipartitions $0:(1,2)$ and $1:(2,0)$, whereas crosstalk errors in the bipartition $2:(1,0)$ are around $0.1\%$. This concludes that detectors for qubits labelled $0$ and $1$ share larger crosstalk correlations.

\section{ Crosstalk Errors in Rigetti Aspen-8 }

We have performed QDT for detectors in Rigetti Aspen-8 on Jan. 7 2021. The detectors of the qubits labelled by $0,1,2,3,15$ and $16$ are considered, see Figure \ref{scenario}. Note that these are chosen out of $31$ qubits contained in Rigetti Aspen-8. Two qubits labelled $0$ and $1$ are not directly connected and the qubit labelled $2$ is directly connected to three qubits $1$, $3$ and $15$.

\begin{center}
\begin{table}
\caption{Two-qubit total errors $D_N$ are computed for pairs of qubits.}
\label{table}
\small
\begin{tabular*}{17.7pc}{ | p{30pt} | p{30pt} |p{30pt} |p{30pt}  | p{30pt} |}
\hline
Qubits &  00 & 01 &10 &11 \\
\hline
\hline
        0,1   & 0.2678 & 0.3394 & 0.2492 & 0.3262 \\
        0,2   & 0.0992 & 0.0960 & 0.0789 & 0.0603 \\
        0,3   & 0.1823 & 0.1512 & 0.1406 & 0.0826 \\
        0,15  & 0.2657 & 0.0881 & 0.3058 & 0.1272 \\
        0,16  & 0.1831 & 0.0560 & 0.2165 & 0.0869 \\
        1,2   & 0.1820 & 0.1961 & 0.3449 & 0.3704 \\
        1,3   & 0.2474 & 0.3022 & 0.4028 & 0.4119 \\
        1,15  & 0.3052 & 0.1651 & 0.4344 & 0.3102 \\
        1,16  & 0.3295 & 0.9706 & 0.9965 & 0.3485 \\
        2,3   & 0.1212 & 0.2107 & 0.2207 & 0.2654 \\
        2,15  & 0.2804 & 0.1614 & 0.2491 & 0.1238 \\
        2,16  & 0.1635 & 0.9910 & 0.9997 & 0.0689 \\
        3,15  & 0.3325 & 0.1152 & 0.3598 & 0.1656 \\
        3,16  & 0.2375 & 0.9899 & 0.9897 & 0.1237 \\
        15,16 & 0.2583 & 0.9946 & 0.9979 & 0.0544 \\
\hline
\multicolumn{5}{@{}p{17pc}@{}} { Large values are found when pairs contain the qubit labelled by $16$.  From the analysis of crosstalk errors in Table \ref{tab7}, the large values are mainly due to local noise. }
\end{tabular*}
\label{tab6}
\end{table}
\end{center}

In Table \ref{tab6}, two-qubit total errors are computed for all of the pairs of qubits. For the qubit $0$, that is not directly connected to the rest of qubits, show total errors $6$-$32\%$. It is found that the pair $(0,2)$ contain the smallest total error. The qubit $2$, connected to three qubits, shows errors $12$-$99\%$. Large values of the errors are found for the pairs containing the qubit labelled $16$. As it is discussed, the total errors can be decomposed into crosstalk and local errors.

The source of the errors can be noticed from the result of computing crosstalk errors, which is shown in Table \ref{tab7}. We first note that the crosstalk errors are not peaked at the pairs containing qubit $16$. This concludes that large values in two-qubit total errors in Table \ref{tab6} are due to local noise. The qubit $0$, connected to none of the rest of qubits, and the qubit $2$, connected to three qubits, do not show much difference in the crosstalk errors. Overall, two-qubit detectors contain relatively higher values about $10$-$40\%$ in the total errors, in which crosstalk errors are around $2$-$7\%$.

\begin{center}
\begin{table}
\caption{  Two-qubit crosstalk errors $D_C$ are computed for pairs of qubits. }
\label{table}
\small
\begin{tabular*}{17.7pc}{ | p{30pt} | p{30pt} |p{30pt} |p{30pt}  | p{30pt} |}
\hline
Qubits &  00 & 01 &10 &11 \\
\hline
\hline
        0,1   & 0.0587 & 0.0485 & 0.0622 & 0.0472 \\
        0,2   & 0.0299 & 0.0244 & 0.0339 & 0.0259 \\
        0,3   & 0.0293 & 0.0288 & 0.0302 & 0.0264 \\
        0,15  & 0.0198 & 0.0316 & 0.0378 & 0.0348 \\
        0,16  & 0.0262 & 0.0192 & 0.0183 & 0.0285 \\
        1,2   & 0.0484 & 0.0715 & 0.0478 & 0.0265 \\
        1,3   & 0.0684 & 0.1050 & 0.0912 & 0.0514 \\
        1,15  & 0.0469 & 0.0484 & 0.0448 & 0.0279 \\
        1,16  & 0.0642 & 0.0590 & 0.0735 & 0.0633 \\
        2,3   & 0.0317 & 0.0655 & 0.0409 & 0.0515 \\
        2,15  & 0.0319 & 0.0332 & 0.0235 & 0.0338 \\
        2,16  & 0.0301 & 0.0328 & 0.0168 & 0.0203 \\
        3,15  & 0.0714 & 0.0584 & 0.0512 & 0.0722 \\
        3,16  & 0.0660 & 0.0519 & 0.0498 & 0.0220 \\
        15,16 & 0.0278 & 0.0432 & 0.0262 & 0.0248 \\
\hline
\multicolumn{5}{@{}p{17pc}@{}} { Crosstalk errors vary from $2$ to $10\%$, which amount to two-qubit gate or measurement errors. }
\end{tabular*}
\label{tab7}
\end{table}
\end{center}

\section{  Entanglement in Measurement Crosstalk Errors } 
Crosstalk errors in a quantum measurement can be rephrased as correlations shared by detectors. The correlations may be investigated at the angle of entanglement theory that characterizes quantum correlations which do not have a classical counterpart. In what follows, we apply entanglement theory to the analysis of correlations in multi-qubit POVM elements of IBMQ Yorktown and Rigetti Aspen-8.

\subsection{Two-qubit POVM elements }
Two-qubit POVM elements correspond to $4\times 4$ non-negative operators. We call a normalized two-qubit POVM element contains entanglement if it cannot be decomposed into a probabilistic mixture of local POVM elements. Otherwise, the POVM element is called separable such that it can be prepared by local operations and classical communication:
\bea
\widetilde{\Pi}_{ab} = \sum_{\lambda} p({\lambda} ) \widetilde{\Pi}_{a}(\lambda) \otimes \widetilde{\Pi}_{b} (\lambda) \nonumber
\eea
for some $\lambda$. For the two-qubit cases, a simple condition can tell whether entanglement is contained or not. A normalized two-qubit POVM is entangled if it is non-positive under the partial transpose (NPPT), where the partial transpose may be taken either of the qubits \cite{ppt}. Otherwise, it contains correlations that can be prepared by local operations and classical communication. 

All two-qubit POVM elements obtained from IBMQ Yorktown in Table \ref{tab4} and Rigetti Aspen-8 in Table \ref{tab6} are analyzed. We have found that none of the pairs is NPPT. This shows that two-qubit crosstalk errors are from classical correlations shared by two detectors. 

\begin{center}
\begin{table*}
\caption{ Entanglement and crosstalk errors in IBMQ Yorktown}
\label{table5}
\small
\begin{tabular*}{35.2 pc}{ | p{70pt} |p{30pt} |p{30pt} |p{30pt}  | p{30pt} | p{30pt} |p{30pt} |p{30pt} |p{30pt} | }
\hline
Outcomes &  000 & 100 & 010 & 110 & 001 & 101 & 011 & 111 \\
	\hline
	\hline
	(0,1,2) & (N,N,N) & (P,P,P) & (P,P,N) & (P,P,P) & (N,N,N)  & (N,N,P) & (N,N,N) & (P,N,P) \\
\hline
\multicolumn{9}{@{}p{35.2pc}@{}} {  For three qubits $(0,1,2)$, bipartite splittings $(\Gamma_{A:BC},\Gamma_{B:CA} ,\Gamma_{C:AB})$ are considered. NPPT correlations are denoted by N, in which entanglement is contained. Otherwise, it is written P. }
\end{tabular*}
\label{tab8}
\end{table*}
\end{center}

\subsection{Three-qubit POVM elements }

Correlations shared by three qubits can be approached by bipartite splittings. It is clear that entanglement exists if correlations in a bipartition are NPPT. For instance, for a detector for three qubits $ABC$, the partial transpose is performed in its normalized POVM element in the bipartite splitting $A:BC$. Negative eigenvalues conclude entanglement in the cut. The partial transpose is applied to other bipartite splittings, $B:CA$ and $C:AB$. 

Contrast to two-qubit cases, we have found that entanglement exists in three-qubit POVM elements. In Table \ref{tab8}, NPPT correlations in IBMQ Yorktown are summarized. In particular, POVM elements giving outcomes $000$, $001$, and $011$ are NPPT in all bipartite splittings. In the case of Rigetti Aspen-8, four triples of qubits $(0,1,2)$, $(0,1,3)$, $(0,1,15)$, and $(1,2,16)$ are considered. It is found that all bipartite splittings of all of the triples contain NPPT correlations. 

\begin{figure*}
\centerline{\includegraphics[width=37pc]{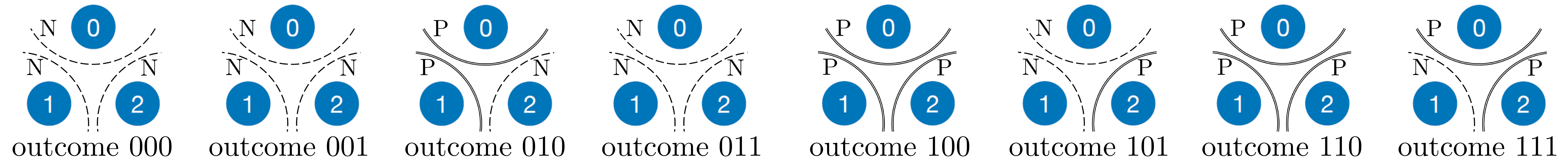}}
\caption{ For qubits labeled $0,1,2$, POVM elements for $i,j,k\in\{0,1\}$ contain entanglement. $N$ stands for NPPT meaning entanglement in the bipartite splitting. Entanglement is detected in all POVM elements except the cases $100$ and $110$, see also Table \ref{table5}.   }
\label{3qubit}
\end{figure*}

\subsection{The structure of correlations }

To summarize, it is found that three-qubit POVM elements contain entanglement whereas two-qubit POVM elements are separable. To be precise, two-qubit POVM elements marginal from their three-qubit extension are separable where the three-qubit POVM elements are entangled. Note also that three-qubit POVM elements contain NPPT correlations depending on a bipartition. The analysis shows that correlations among detectors are highly non-trivial. For instance, it is necessary to perform QDT for more than two qubits to understand the correlation structure. 

\section{Application to Quantum Error Mitigation}

Let $M_{abc}^{ABC}$ denote a POVM element of three qubits labeled $ABC$ giving outcome $abc$. The probability of outcome $abc$ is given by
\bea
p_{exp}(abc) = \tr[ \rho^{A}\otimes \rho^{B}\otimes \rho^C M_{abc}^{ABC}]. \nonumber
\eea
for a three-qubit state $\rho^{A}\otimes \rho^{B}\otimes \rho^C$. Let $p_0(abc)$ denote the probability with a POVM element $|a\rangle \langle a| \otimes |b\rangle\langle b|\otimes |c\rangle \langle c|$ in a noiseless environment. In error mitigation, it is aimed to transform a probability given from an experiment, $p_{exp}(abc)$, to a probability as close to the ideal one $p_{0}(abc)$. When noise is present in qubits individually, i.e., with $M_{abc}^{ABC} = M_{a}^A\otimes M_{b}^{B}\otimes M_{c}^C$ for some POVM elements $M_{a}^A$, $M_{b}^{B}$ and $M_{c}^C$, the probability can be factorized for individual qubits. Thus, errors can be mitigated by taking individual qubits only into account \cite{HH1, II1}. This strategy no longer works when crosstalk errors exist. Hence, our results show that measurement errors on multiple qubits can be mitigated by considering outcomes of multiple qubits collectively.

\section{Conclusion}

In conclusion, we have analyzed measurement readout errors in cloud-based quantum computing services. A method of quantifying crosstalk errors is introduced and applied to analyzing readout errors in IBMQ Yorktown and Rigetti Aspen-8. Our results characterize and quantify measurement crosstalk errors in quantum computing with NISQ technologies, and pave a way to devise a method of dealing with measurement readout errors in multiple qubits, such as measurement error correction or mitigation. Our results also suggest that QDT may be performed before qubit allocations so that the crosstalk errors are analyzed and those qubits having a large fraction of crosstalk in a measurement are avoided by the command of qubit allocations. It would be interesting to investigate the relations of error mitigation and the properties and the structure of crosstalk errors in a quantum measurement. 

Finally, we address that raw data from QDT were generated at the cloud-based quantum computing services IBMQ and Rigetti. Derived data supporting the findings of this study are available from the corresponding author upon request.

\section{Acknowledgement}
This work was supported by Samsung Research Funding $\&$ Incubation Center of Samsung Electronics (Project No. SRFC-TF2003-01).

\end{document}